\documentclass[aps,prl,showpacs,amssymb,superscriptaddress,footinbib,twocolumn]{revtex4}
\usepackage[ansinew]{inputenc}
\usepackage{bbm}
\usepackage{bm}
\usepackage{amsbsy}
\usepackage{amsthm}
\usepackage{amssymb}
\usepackage{amsfonts}
\usepackage{amsmath}
\usepackage{dsfont} 
\usepackage{graphicx} 
\usepackage{epsfig}
\usepackage{epstopdf}
\usepackage{dsfont}
\usepackage{color}
\usepackage[colorlinks]{hyperref}
\usepackage[figure,table]{hypcap}
\hypersetup{
	bookmarksnumbered,
	pdfstartview={FitH},
	citecolor={darkgreen},
	linkcolor={darkred},
	urlcolor={darkblue},
	pdfpagemode={UseOutlines}}
\definecolor{darkgreen}{RGB}{50,190,50}
\definecolor{darkblue}{RGB}{0,0,190}
\definecolor{darkred}{RGB}{238,0,0}
\usepackage{soul}
\newcommand{\pr}{^{\prime}}

\DeclareMathOperator{\diag}{diag}

\newcommand{\tr}{\textnormal{Tr}}

\begin{document}

\title{Relativistic Quantum Teleportation with Superconducting Circuits}

 \author{N. Friis}
\affiliation{School of Mathematical Sciences,
University of Nottingham,
University Park,
Nottingham NG7 2RD,
United Kingdom}

\author{A. R. Lee}
\affiliation{School of Mathematical Sciences,
University of Nottingham,
University Park,
Nottingham NG7 2RD,
United Kingdom}

\author{K. Truong}
\affiliation{School of Mathematical Sciences,
University of Nottingham,
University Park,
Nottingham NG7 2RD,
United Kingdom}

\author{C. Sab\'in}
\affiliation{School of Mathematical Sciences,
University of Nottingham,
University Park,
Nottingham NG7 2RD,
United Kingdom}

\author{E. Solano}
\affiliation{Departamento de Qu\'{\i}mica F\'{\i}sica, Universidad del Pa\'{\i}s Vasco UPV/EHU, Apartado\ 644, 48080 Bilbao, Spain}
\affiliation{IKERBASQUE, Basque Foundation for Science, Alameda Urquijo 36, 48011 Bilbao, Spain}

\author{G. Johansson}
\affiliation{Microtechnology and Nanoscience, MC2, Chalmers University of Technology,
S-41296, G\"{o}teborg, Sweden}

\author{I. Fuentes}
 \affiliation{School of Mathematical Sciences,
University of Nottingham,
University Park,
Nottingham NG7 2RD,
United Kingdom}
\date{\today}

\begin{abstract}
 We study the effects of relativistic motion on quantum teleportation and propose a realizable experiment where our results can be tested. We compute bounds on the optimal fidelity of teleportation when one of the observers undergoes nonuniform motion for a finite time. The upper bound to the optimal fidelity is degraded due to the observer's motion. However, we discuss how this degradation can be corrected. These effects are observable for experimental parameters that are within reach of cutting-edge superconducting technology. Our setup will further provide guidance for future space-based experiments.

\end{abstract}

\maketitle
\emph{How are quantum information tasks affected by relativistic motion?} This seemingly simple question has been intriguing researchers for more than a decade~\cite{PeresTerno2004} and it has been the cornerstone of an entire new field of physics, \emph{relativistic quantum information}; see, e.g., Ref.~\cite{AlsingFuentes2012} for a recent review. In addition to its theoretical interest, the topic is increasingly gaining practical relevance as quantum information experiments are reaching relativistic regimes~\cite{RideoutEtal2012}. However, a satisfactory, empirically testable framework to address this question has been missing.

The first attempts to answer this open question considered highly idealized situations where observers with constant, eternal accelerations analyzed the entanglement between global modes of a quantum field~\cite{AlsingMilburn2003,Fuentes-SchullerMann2005,BruschiLoukoMartin-MartinezDraganFuentes2010}. Typically, these studies consider only the mathematical intricacies of the problem and contain little reference to realistic physical setups. However, recent theoretical work~\cite{BruschiDraganLeeFuentesLouko2012,DownesFuentesRalph2011} has analysed relativistic entanglement in paradigmatic quantum optical scenarios such as cavity QED~\cite{RaimondBruneHaroche2001}. In the relativistic case cavities move with accelerations that can arbitrarily vary in time.  Interestingly, these works also show that relativistic motion can also be used to implement quantum gates~\cite{BruschiDraganLeeFuentesLouko2012}.

On the other hand, the \emph{dynamical Casimir effect}, a phenomenon that involves extreme accelerations and relativistic velocities, has recently been demonstrated in a real experiment using superconducting circuits where the relativistic motion of an effective boundary condition was successfully implemented~\cite{WilsonDynCasNature2012}.

\begin{figure}[t!]
\includegraphics[width=0.8\columnwidth]{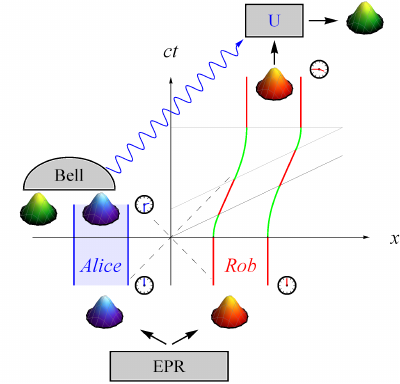}
\caption{Alice and Rob initially share a two-mode squeezed cavity state that is produced by the EPR source. Consecutively, Rob's cavity undergoes nonuniform motion consisting of segments of constant acceleration (green hyperbolae) and inertial coasting (red parallel lines). Alice, who remains inertial, sends the outcome of the Bell measurement on the input state and her mode of the entangled
state to Rob via a classical channel (blue wavy arrow). Rob can then retrieve the teleported state by performing the appropriate unitary $U$. In addition to the standard protocol, both Alice and Rob measure their respective proper times and perform local rotations to compensate for the phases accumulated during the motion.}
\label{fig:teleportation in motion}
\end{figure}

In this Letter we show that nonuniformly accelerated motion has  effects on a paradigmatic quantum information protocol,
\emph{quantum teleportation}, in a framework in which the theoretical predictions can be  tested. We focus on the effects of nonuniform
motion on the fidelity of the standard protocol for quantum teleportation with continuous variables; see Fig.~\ref{fig:teleportation in motion}. We employ
the powerful tools of quantum optics for Gaussian states and apply them in the microwave regime. In this setting we can take advantage
of very recent experimental developments in circuit quantum electrodynamics~\cite{YouNori2011}.

Our main results are the following. We observe two distinguishable degradation effects.  The first is due to the time evolution of the field. The fidelity loss due to this effect can be corrected by applying local operations which depend on the proper time. The second degradation effect is solely due to nonuniform acceleration of the rigid cavity and it can only be avoided by conveniently choosing the duration of the accelerated part of the motion. The latter effect originates from the particle generation caused by Rob's motion, which entangles the modes of his cavity. Consequently, due to the monogamy of entanglement, the original entanglement between the modes selected by Alice and Rob is degraded. This mechanism also lies at the heart of the dynamical Casimir effect and the Unruh-Hawking effect. Furthermore, the model we use to describe nonuniform motion of rigid cavities directly applies also to quantum optical equipment incorporated in satellites and will therefore be relevant for quantum communication tasks and high-sensitivity tests of quantum optics in space-based experiments.

\begin{figure}[t!]
\includegraphics[width=0.85\columnwidth]{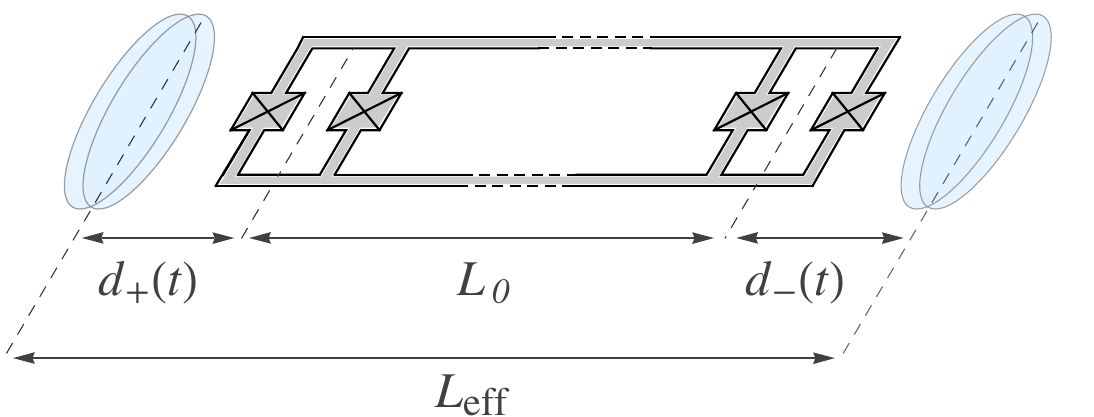}
\caption{\textbf{Sketch of the experimental setup}. A coplanar waveguide is interrupted by two \emph{superconducting quantum interference devices}
(SQUIDs) placed at a fixed distance $L_{0}$, creating a cavity of effective length $L_{\mathrm{eff}}=L_{0}+d_{+}(t)+d_{-}(t)$. The
time dependence of the distances $d_{\pm}(t)$ between the SQUIDs and the effective boundaries is controlled with external drive fields applied to
the superconducting circuits to simulate a cavity of constant length with respect to its rest frame.}
\label{fig:microwave cavity setup}
\end{figure}

Finally, we introduce our experimental setup, see Fig.~\ref{fig:microwave cavity setup}, to test our results using cutting-edge circuit QED technology. We generalize the idea~\cite{WilsonDynCasNature2012} of producing a single oscillating boundary condition and now describe the case of a rigid cavity moving with constant acceleration during a finite time interval, that is, two boundary conditions moving in a coordinated fashion.

Let us now discuss our model, starting with the standard teleportation protocol in continuous variable systems which assumes Alice and Bob to be at rest at all times~\cite{MariVitali2008}. The novelty of our approach will be to consider that Rob, the relativistic Bob, undergoes nonuniform, relativistic motion before concluding the protocol. Initially, Alice and Rob are at rest and share a two-mode squeezed state of a $(1+1)$-dim cavity field with squeezing parameter $r>0$. The quantum correlations of this state are characterized by its covariance matrix $\sigma_{kk\pr}=\begin{pmatrix} A & C\, \\ C^{T} & B\, \end{pmatrix}$, where $A$, $B$, and $C$ are real $2\times2$ matrices that only depend on $r$; see, e.g., Ref.~\cite{AdessoIlluminati2005}. We label Alice's and Rob's modes by $k$ and $k\pr$, respectively.
Alice wants to use the entanglement of this state as a resource to teleport an additional unknown coherent state to Rob. More precisely, Alice wants to teleport the information contained in the vector of first moments. The teleportation of the full state is, in fact, impossible; see, e.g., Ref.~\cite{LaihoMolotkovNazin2000}. To this end she performs a beam-splitting operation on her mode $k$ and the input state. Subsequently, she performs a homodyne measurement, and communicates the results to Rob via a classical channel. After receiving Alice's measurement outcome, Rob performs the according displacement operation on his mode $k\pr$ in order to recover Alice's coherent state. Given the covariance matrix $\sigma_{kk\pr}$, the fidelity of this protocol when Rob remains at rest is
\begin{equation}
\mathcal{F} = \dfrac{2}{\sqrt{4+\tr(N)+ 2\det(N)}}\,,
\label{eq:fidelity}
\end{equation}
where $N=ZAZ+ZC+C^{T}Z+B$ and $Z=\diag(1,-1)$. In addition, Alice and Rob can use local operations and classical communication to improve the fidelity of the protocol without increasing the amount of shared entanglement. In particular, optimizing over all Gaussian local operations~\cite{MariVitali2008} the upper bound to the optimal fidelity of teleportation can be expressed as
\begin{equation}
\mathcal{F}_{\mathrm{opt}} \leq \dfrac{1}{1+\nu}\,,
\label{eq:optimalfidelity}
\end{equation}
where $\nu$ is the so-called ``smallest symplectic eigenvalue'' of the partial transpose of $\sigma_{kk\pr}$. Since an infinite amount of squeezing
cannot be produced, perfect teleportation is not possible, but the upper bound is achieved precisely if Alice and Rob share a two-mode squeezed state,
for which $\nu=\exp(-2r)$, $\infty>r>0$. Then $\mathcal{F}=\mathcal{F}_{\mathrm{opt}}=1/\bigl[1+\exp(-2r)\bigr]$. As we have not yet taken into account the observers' motion, it comes as no surprise that the optimal fidelity of teleportation only depends on the squeezing, that is, the amount of initial entanglement.
\begin{figure*}[t]
\hspace*{-0.4cm}
\includegraphics[width=0.5\linewidth]{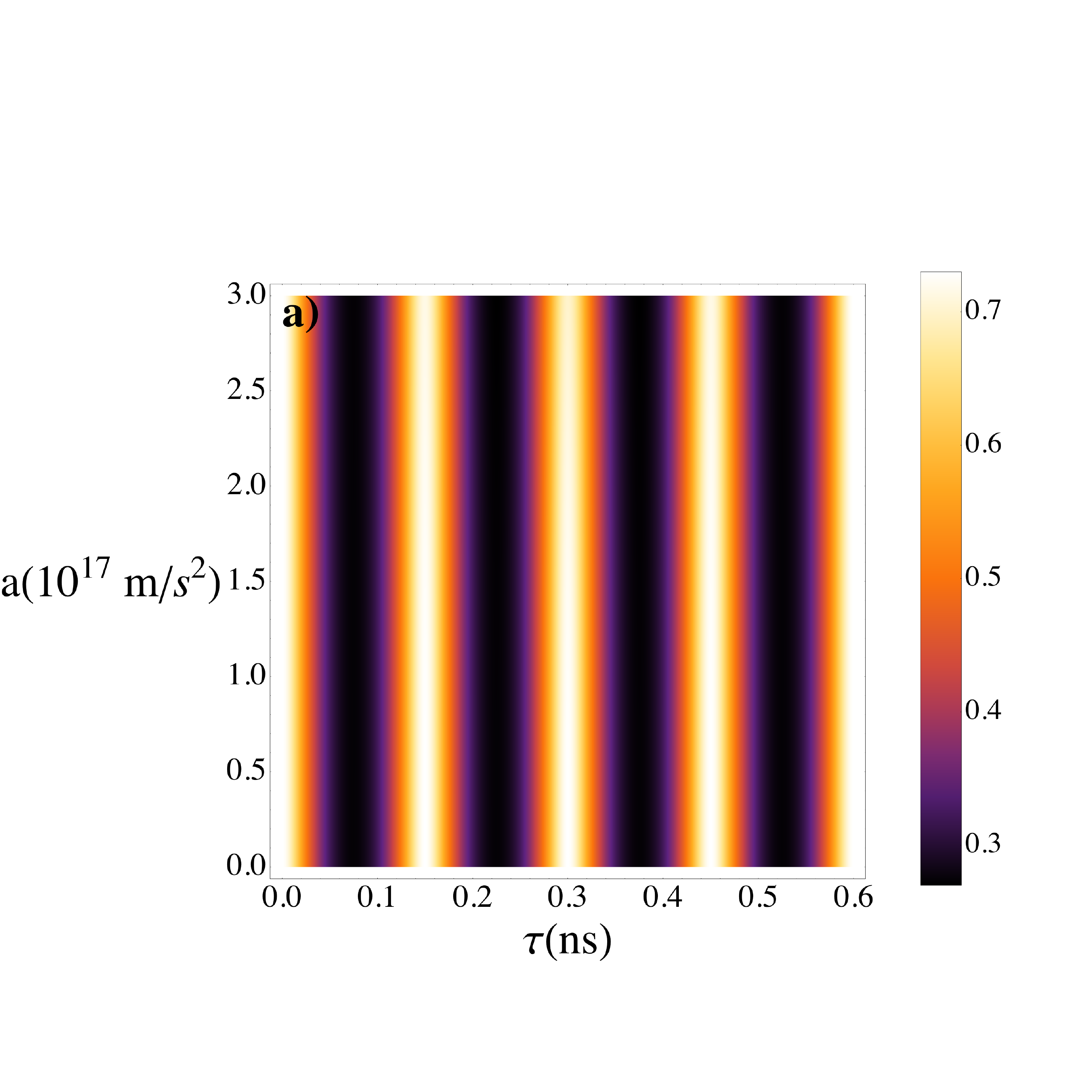}
\includegraphics[width=0.5\linewidth]{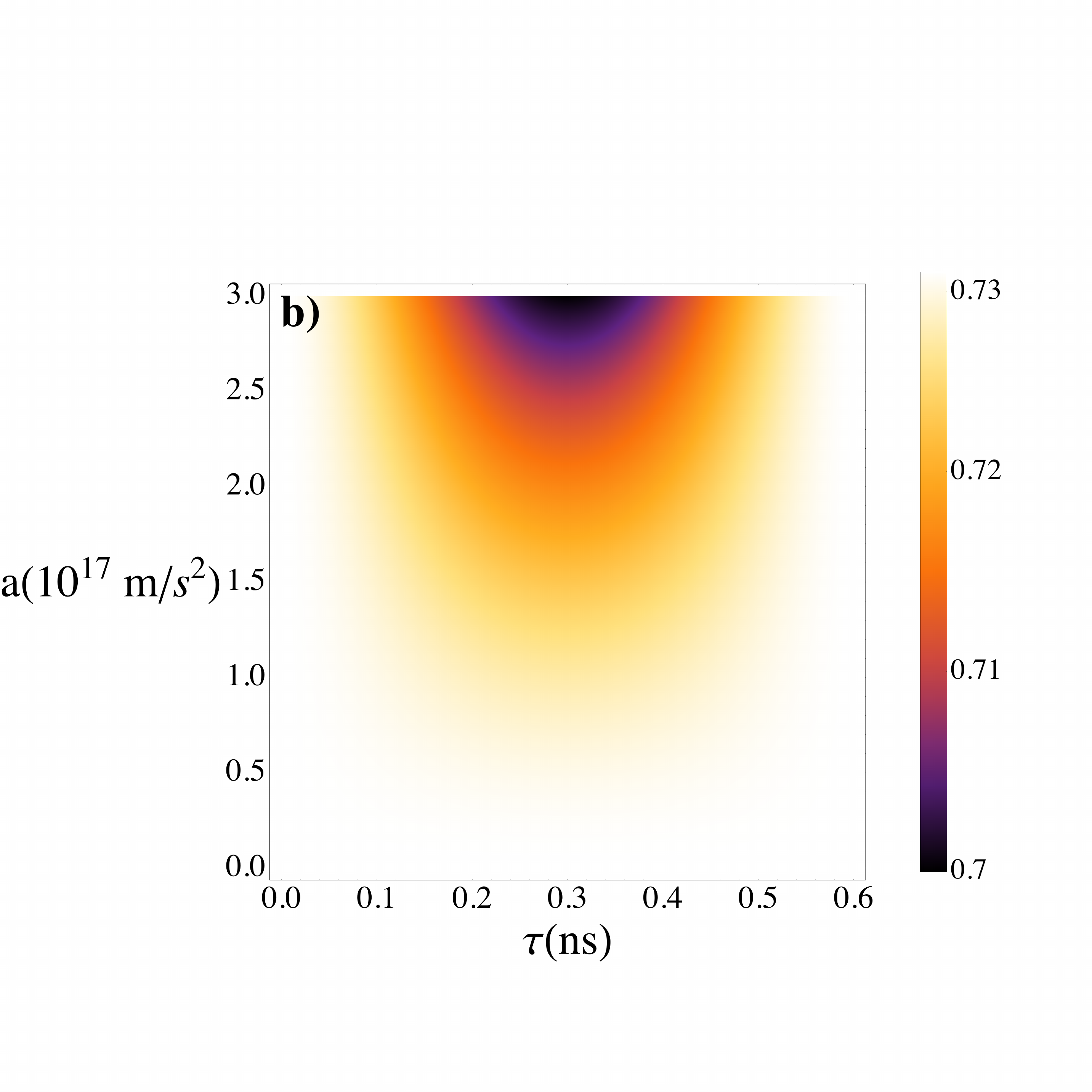}
\caption{The fidelities $\tilde{\mathcal{F}}$ and $\tilde{\mathcal{F}}_{\mathrm{opt}}$ are plotted in~(a) and~(b), respectively, as functions of Rob's proper time $\tau$ and the acceleration $a$. The plots are shown for modes $k=1$ (Alice) and $k\pr=3$ (Rob). For the cavity length we use a typical value of $L=1.2\,\operatorname{cm}$. Together with the speed of light $c=1.2\times10^8\,\mathrm{m/s}$ we obtain a fundamental frequency $\omega_{k}=2\pi\times c/(2L)=2\pi\times 5\mathrm{GHz}$ and $\omega_{k\pr}=3\omega_{k}$. The squeezing parameter is $r=1/2$ and the maximum value of the perturbative parameter is $h^{2}=0.06$. Once the effect of the free time evolution in~(a) is removed, the correction due to acceleration can be isolated in~(b).}
\label{fig:results}
\end{figure*}

Now let us consider the scenario that is sketched in Fig.~\ref{fig:teleportation in motion}.
After the preparation of the initial state, Rob's cavity undergoes nonuniform motion, consisting of periods of constant acceleration and inertial motion. Because of the motion, the covariance matrix $\sigma$  is transformed to $\tilde{\sigma}=S\sigma S^{T}$, where $S$ is a symplectic matrix. The reduced covariance matrix $\tilde{\sigma}_{kk\pr}$ for two modes $k$ and $k\pr$ can be obtained from $\tilde{\sigma}$.
If the motion is inertial for the time $t$, then $S$ is simply composed of local rotations with angles $\omega_{k}t$ and $\omega_{k\pr}t$, where $\omega_{k}$ and $\omega_{k\pr}$ are the angular frequencies of the modes $k$ and $k\pr$, respectively. We let Rob's cavity accelerate for a proper time $\tau$ which is measured at the center of the cavity. Then $S$ is given in terms of the Bogoliubov coefficients $\alpha_{mn}$ and $\beta_{mn}$ that relate the mode functions of the inertial and accelerated cavity~\cite{FriisFuentes2012}. The coefficients $\alpha_{mn}$ account for mode mixing, while $\beta_{mn}$ account for particle pair production. The coefficients can be computed analytically using a perturbative expansion in the parameter
\begin{equation}
h=
a\,L/c^{2}\ll 1\,,
\label{eq:pertparam}
\end{equation}
where $a$ is the acceleration at the centre of the cavity, $L$ is the cavity's length at rest, and $c$ is the propagation speed of the field excitations in the cavity. We can write the coefficients as $\alpha_{mn}=\alpha^{_{(0)}}_{mn}+\alpha^{_{(1)}}_{mn}h+O(h^{2})$ and $\beta_{mn}=\beta^{_{(1)}}_{mn}h+\beta^{_{(2)}}_{mn}h^{2}+O(h^{3})$. Notice that, for cavities of typical sizes, this approach can accommodate extremely large accelerations, as we will see in detail below. Moreover, there are no restrictions on the duration, covered distance, or the achieved velocity of the motion. Technical details on this formalism\textemdash involving the quantization of Klein-Gordon fields with cavity boundary conditions and suitable coordinates for inertial and accelerated observers\textemdash can be found, for instance, in~\cite{FriisFuentes2012}. Now we will focus on the effect of the motion on the fidelity of the teleportation protocol described above. Although the formalism allows us to consider arbitrary trajectories, here we consider the simplest case: Rob's motion is inertial apart from one finite interval of constant acceleration.

First, inserting the motion-transformed covariance matrix $\tilde{\sigma}_{kk'}$ into Eq.~(\ref{eq:fidelity}), we find the perturbative expansion of the teleportation fidelity, i.e.,
\begin{align}
\tilde{\mathcal{F}}  &=  \tilde{\mathcal{F}}^{\raisebox{-1.7pt}{\scriptsize{$(0)$}}}- \tilde{\mathcal{F}}^{\raisebox{-1.7pt}{\scriptsize{$(2)$}}}\,h^{2}+\mathcal{O}(h^{4})\,,
\label{eq:fidelity series expansion}
\end{align}
where the expansion coefficients are given by
\begin{subequations}
\label{eq:fidelity series expansion terms}
\begin{align}
\tilde{\mathcal{F}}^{\raisebox{-1.7pt}{\scriptsize{$(0)$}}}    &=
\bigl[1+\operatorname{cosh}(2 r)- \operatorname{cos}(\phi)\operatorname{sinh}(2r)\bigr]^{-1}\,,
\label{eq:fidelity series expansion terms h0}\\[1mm]
\tilde{\mathcal{F}}^{\raisebox{-1.7pt}{\scriptsize{$(2)$}}}    &=
\bigl(\tilde{\mathcal{F}}^{\raisebox{-1.7pt}{\scriptsize{$(0)$}}}\bigr)^{2} \bigl(1+e^{-2r}\bigr)\bigl[f^{\beta}_{k\pr}+f^{\alpha}_{k\pr}\,\operatorname{tanh}(2r)\bigr]\,,
\label{eq:fidelity series expansion terms h2}
\end{align}
\end{subequations}
and $\phi=\omega_{k}t+\omega_{k\pr}\tau$. The additional expressions $f^{\alpha}_{k\pr}=\tfrac{1}{2}\sum_{n}|\alpha_{nk\pr}^{\raisebox{0.7pt}{\tiny{$\,(1)$}}}|^{2}$ and $f^{\beta}_{k\pr}=\tfrac{1}{2}\sum_{n}|\beta_{nk\pr}^{\raisebox{0.7pt}{\tiny{$\,(1)$}}}|^{2}$ in Eq.~(\ref{eq:fidelity series expansion terms h2}) also depend on $\tau$; see Ref.~\cite{FriisFuentes2012}.
Because of the dependence of $\phi$ on Alice's and Rob's proper times, there is a degradation effect on the fidelity as shown in Fig.~\ref{fig:results}(a).
In particular, this also occurs in the inertial case since the free evolution continuously rotates Alice's and Rob's modes which affects the optimal performance of the protocol. Other effects of inertial motion on entanglement have been discussed in several works~\cite{SaldanhaVedral2012a,SaldanhaVedral2012b,PeresScudoTerno2002,GingrichAdami2002,PachosSolano2003, LamataMartinDelgadoSolano2006,FriisBertlmannHuberHiesmayr2010}, but these are not of concern for the cavity modes we consider here. However, the fidelity of teleportation does not solely depend on the entanglement. To correct the effects due to the time dependence of the phase, Alice and Rob must apply local rotations which depend only on their local proper times or choose times such that $\phi=2\pi n$.  In the case $h=0$ the maximal, optimal fidelity $\mathcal{F}_{\mathrm{corr}}=\mathcal{F}_{\mathrm{opt}}=1/\bigl[1+\exp(-2r)\bigr]$ can be recovered.

Remarkably, if the acceleration is nonzero the dependence of $\tilde{\mathcal{F}}^{\raisebox{-1.7pt}{\scriptsize{$(2)$}}}$ on the local phase $\phi$ can be removed by exactly the same local rotations as in the inertial case.  The degradation of the fidelity $\tilde{\mathcal{F}}_{\mathrm{corr}}$ can then be attributed solely to the nonuniform acceleration: the creation of particles entangles Rob's mode with the new modes in his cavity, and due to well-known results on monogamy of entanglement, this leads to a degradation of the original quantum correlations with Alice's mode and thus a loss in teleportation fidelity. Moreover, the protocol including the local phase rotations turns out to be optimal. In other words, the motion-transformed version of Eq.~(\ref{eq:optimalfidelity}), i.e., the upper bound on the fidelity, is achieved by $\tilde{\mathcal{F}}_{\mathrm{corr}}$. In that case, we have $\tilde{\mathcal{F}}_{\mathrm{corr}}=\tilde{\mathcal{F}}_{\mathrm{opt}}=
\tilde{\mathcal{F}}^{\raisebox{-1.9pt}{\tiny{\,$(0)$}}}_{\mathrm{opt}}-
\tilde{\mathcal{F}}^{\raisebox{-1.9pt}{\tiny{\,$(2)$}}}_{\mathrm{opt}}\,h^2 +\mathcal{O}(h^4)$, where $\tilde{\mathcal{F}}^{\raisebox{-1.9pt}{\tiny{\,$(0)$}}}_{\mathrm{opt}}=\mathcal{F}_{\mathrm{opt}}=1/\bigl[1+\exp(-2r)\bigr]$ is the optimal expression for $h=0$ above and
\begin{align}
\tilde{\mathcal{F}}^{\raisebox{-1.9pt}{\tiny{\,$(2)$}}}_{\mathrm{opt}}  &=
\tilde{\mathcal{F}}^{\raisebox{-1.9pt}{\tiny{\,$(0)$}}}_{\mathrm{opt}}
\bigl[f^{\beta}_{k\pr}+f^{\alpha}_{k\pr}\,\operatorname{Tanh}(2r)\bigr]\,.
\label{eq:corroptfid}
\end{align}

In Fig.~\ref{fig:results}(b) we plot $\tilde{\mathcal{F}}_{\mathrm{opt}}$, allowing us to identify a regime of strength and duration of acceleration at which the corrections due to motion amount to $4 \%$ of the total fidelity. As we explain below, this regime is well within experimental reach with current technology. Furthermore, the effect can be amplified by selecting more complicated trajectories~\cite{BruschiDraganLeeFuentesLouko2012}.

Let us now discuss the details of the experimental setup that we propose to test our predictions. We will use state-of-the-art technology in circuit  quantum electrodynamics. Two-mode squeezed states in the microwave regime have been produced in the laboratory with squeezing parameter $r=\operatorname{log}2/2$; see Refs.~\cite{EichlerEtal2011,FlurinRochMalletDevoretHuard2012,MenzelEtal2012}. Beam splitters for propagating photons with frequencies around $5~\operatorname{GHz}$ based on superconducting circuit architectures are also available~\cite{HoffmannEtal2010}. Therefore, we believe that the standard continuous variables teleportation protocol may be realized experimentally. Obviously the most demanding aspect of our proposal is the implementation of highly accelerated motion as a new part of the protocol. To this end we will take advantage of the technology developed for the experiment verifying the dynamical Casimir effect~\cite{WilsonDynCasNature2012}. While in \cite{WilsonDynCasNature2012} the boundary conditions are equivalent to a single mirror oscillating in free space, we are interested in cavities undergoing nonuniform accelerated motion.

The cavities of our setting can be engineered as a coplanar microwave waveguide terminated by two \mbox{dc SQUIDs} placed at a distance $L_{0}$ from each other~\cite{SvenssonMScThesis2012}; see Fig.~\ref{fig:microwave cavity setup}. The SQUIDs generate boundary conditions for the (1+1)-dim quantum field along the waveguide~\cite{JohanssonJohanssonWilsonNori2010}, producing a cavity of constant effective length $L_{\mathrm{eff}}$ with respect to its rest frame. The boundary condition depends on the external magnetic flux threading the SQUID. In Rob's case, the time variation of this flux amounts to a time variation of $d_{\pm}$ to produce the different  effective accelerations of the boundaries, which will keep the cavity length fixed in its rest frame. Therefore, by applying external drive fields on both SQUIDS with appropriate time profiles, the system becomes equivalent to a cavity of constant length in motion. This setup has already been implemented in the laboratory~\cite{SvenssonMScThesis2012} and the cavity accommodates a few modes below the natural cutoff provided by the plasma frequency of the SQUID. In contrast to the oscillating motion of~\cite{SvenssonMScThesis2012}, the profile of the driving fields can be adjusted to mimic constant accelerated motion during a finite interval of time. Taking as a reference the oscillating motion of a single mirror in~\cite{WilsonDynCasNature2012} with a driving frequency $\omega_{D}= 2\pi\times 10\, \mathrm{GHz}$ and an amplitude of $0.1\,\mathrm{mm}$ for the effective motion, the maximum acceleration achieved was $4\times10^{17}\,m/s^{2}$. These realistic values are enough to observe our predictions, since they give rise to values of $h$ larger than the ones considered in Fig.~\ref{fig:results}(b). Other sources of experimental degradation of the fidelity, like losses in the beam splitters and inefficiencies of the detectors, would only reduce the maximum fidelity attainable but would exhibit a qualitatively different dependence on the magnitude and duration of the acceleration.

To summarize our results, we have analyzed the effect of relativistic motion on the fidelity of the standard continuous variable protocol for quantum teleportation. The effects of nonuniform acceleration on the  fidelity can be isolated by applying proper-time dependent local operations which remove the effects of time evolution. We have shown that the degradation of the fidelity due to acceleration is sizeable for realistic experimental parameters. We have further suggested a particular experimental setup with superconducting cavities that is well within reach of state-of-the-art technology.  The origin of the fidelity loss is the same physical mechanism\textemdash particle generation due to motion\textemdash underlying the dynamical Casimir effect and the Unruh-Hawking radiation. Therefore, its observation would also shed light on these phenomena. Moreover, via the equivalence principle, our results suggest the existence of observable effects of gravity on quantum information setups, which may be relevant for space-based experiments~\cite{RideoutEtal2012}. Finally, it is possible that theoretical predictions derived with a similar formalism, e.g., the implementation of quantum gates by cavity motion~\cite{BruschiDraganLeeFuentesLouko2012}, may be realizable in similar experiments. We believe that the effects studied in relativistic quantum information scenarios will finally leave the realm of theoretical gedanken experiments. The analysis of relativistic effects on quantum information can now be extended by empirical tests. Furthermore, low-cost experimental setups to test relativistic aspects of quantum communication, such as the one proposed here, will inform future space-based, high-risk experiments.

\begin{acknowledgements}
We thank Gerardo Adesso, David Edward Bruschi, Marcus Huber, Jorma Louko, Borja Peropadre and Guillermo Romero for useful discussions and comments. N.~F., C.~S. and I.~F. acknowledge support from EPSRC (CAF Grant No.~EP/G00496X/2 to I.~F.),  K.~T. acknowledges support from EPSRC Vacation Bursary 2012. C.~S. acknowledges the Spanish research consortium QUITEMAD S2009-ESP-1594. G.~J. acknowledges support from the Swedish Research Council and from the EU through the ERC and the FET-OPEN project PROMISCE. E.~S. acknowledges funding from the Spanish MINECO project FIS2012-36673-C03-02; UPV/EHU UFI 11/55; Basque Government IT472-10; SOLID, CCQED, PROMISCE, and SCALEQIT European projects.
\end{acknowledgements}

\end{document}